# Energetics and Kinetics Requirements for Organic Solar Cells to Break the 20% Power Conversion Efficiency Barrier

Oskar J. Sandberg*,[†] and Ardalan Armin*,[†]

[†]Sustainable Advanced Materials (Sêr-SAM), Department of Physics, Swansea University, Singleton Park, Swansea SA2 8PP Wales, United Kingdom

Corresponding Authors

*E-mail: o.j.sandberg@swansea.ac.uk

*E-mail: ardalan.armin@swansea.ac.uk

**Abstract**

The thermodynamic limit for the efficiency of solar cells is predominantly defined by the energy bandgap of the used semiconductor. In case of organic solar cells both energetics and kinetics of three different species play role: excitons, charge transfer states and charge separated states. In this work, we clarify the effect of the relative energetics and kinetics of these species. Making use of detailed balance, we develop an analytical framework describing how the intricate interplay between the different species influence the photocurrent generation, the recombination, and the open-circuit voltage in organic solar cells. We clarify the essential requirements for equilibrium between excitons, CT states and charge carriers to occur. Furthermore, we find that the photovoltaic parameters are not only determined by the relative energy level between the different states but also by the kinetic rate constants, highlighting the importance of slow exciton recombination in at low energetic offsets. Finally, depending on the kinetic parameters, we find an optimal power conversion efficiency exceeding 20% at energetic offsets around 0.1 eV. These findings provide vital insights into the operation of state-of-art non-fullerene organic solar cells with low offsets.

# 1. Introduction

Recently, organic solar cells based on donor-acceptor (D-A) bulk heterojunctions (BHJs) have seen a drastic increase in the device performance, with power conversion efficiencies (PCEs) currently exceeding 17%,[1,2] with 20% in sight and even 25% predicted.[3] This has sparked a renewed interest in photovoltaic applications based on organic semiconductors. The increased efficiency has been achieved by using strongly absorbing narrow-gap non-fullerene acceptors,[3,4] resulting in better complementary light absorption by both the donor and the acceptor and significantly increased short-circuit current densities ($J_{SC}$). Simultaneously, the energetic offset between the donor and the acceptor has been decreased allowing for losses in the open-circuit voltage ($V_{OC}$) to be reduced. The charge generation yield and the photovoltage in organic solar cells is ultimately determined by physical processes taking place between excitons (S), interfacial charge-transfer (CT) states, and separated free charge carriers (CS). However, the interplay between these different species continues to be a matter of controversy.

In general, the conversion of strongly bound excitons, formed in neat donor or acceptor phases under illumination, into free charge carriers (producing electricity) is mediated by CT states, constituting bound electron-hole pairs where the electron is in the acceptor phase and the hole in the donor phase.[5] Compared to excitons, CT states are considerably less bound, dissociating into free charge carriers with relatively high quantum efficiencies, although the underpinning mechanism is still debated.[6-13] Similarly, also the recombination of free charge carriers is believed to take place via CT states. While the encounter between free electrons and holes follows a diffusion-limited Langevin-like process, albeit geometrically constricted,[14] every charge encounter is expected to result in a CT state which may subsequently dissociate back into free charge carriers or recombine.[15-17] This interrelation has been suggested to induce a mutual equilibrium between CT states and free charge carriers in so-called reduced Langevin systems where CT state dissociation is efficient.[18] Such an equilibrium further implies that the $V_{OC}$ in organic solar cells is entirely defined by the energetics and the (radiative and non-radiative) recombination kinetics of CT states. This has indeed been observed in fullerene-acceptor-based BHJs, characterized by a large energy offset between excitons and CT states.[18-21]

Whether a similar situation applies for non-fullerene systems displaying low energy offsets between excitons (in the acceptor) and CT states has remained unclear. Recent studies of the radiation efficiency in state-of-the-art non-fullerene solar cells found these systems to possess low non-radiative losses.[22,23] This has been attributed to the formation of an equilibrium between excitons in the acceptor and CT states, causing a repopulation of the presumably more radiative excitons.[23-26] It has also been pointed out that the small offset between excitons and CT states might compromise the charge generation yield, as the driving force for exciton dissociation is reduced.[27] On the other hand, it has been suggested that if the exciton lifetime is long enough, even in case of low charge transfer rates associated with a small exciton-CT offset, an efficient charge transfer can still occur.[23] However, comprehensive analytical treatments relating the relevant rates and energetics to key performance parameters, such as the $V_{\text{OC}}$, in low-offset systems are still lacking. Moreover, the conditions for when an equilibrium between excitons, CT states and free charge carriers can be established, and how this equilibrium affect the overall device performance, have thus far remained elusive.

In this work, the interplay between free charge carriers, CT states and excitons, and its relation to the device performance in low-offset organic solar cells, is investigated. Based on detailed balance considerations, we derive analytical expressions relating the charge generation yield, the charge-carrier recombination coefficient, and the open-circuit voltage to the relative energetics and kinetics between the different species. Furthermore, we clarify the conditions for when mutual equilibrium between the different states can occur. Finally, we demonstrate how the relative energetics and kinetics critically determine the overall power conversion efficiency in these solar cells, suggesting that a PCEs above 20% may be obtained at energetic offsets of 0.1 eV.

**2. Theory**

We consider a BHJ device with an active layer of thickness $d$, where the (lowest) exciton energy of the donor is much higher than both the (lowest) exciton energy $E_S$ of the acceptor and the CT state energy $E_{\text{CT}}$, such that the back transfer from CT states to donor excitons is negligible. A schematic state diagram showing the relevant rates between the different species is depicted in **Figure 1**. Here, triplet states are deliberately omitted to simplify the analysis.[16] Furthermore, the effect of trap-assisted charge-carrier recombination is assumed to be negligible.[28] Under steady-state illumination conditions, the kinetics between excitons (in the acceptor), CT states and separate free charge carriers are described by the following rate equations:

$$\frac{\partial n_S}{\partial t} = G_S^{(A)} - k_S n_S - k_{ht} n_S + k_{bt} n_{CT} = 0, \qquad (1a)$$

$$\frac{\partial n_{CT}}{\partial t} = G_{CT} + k_{ht} n_S - k_{bt} n_{CT} - k_f n_{CT} - k_d n_{CT} + \beta_0 n_{CS}^2 + \Delta G_{\text{eff}}^{(D)} = 0, \quad (1b)$$

$$\frac{J - J_{\text{surf}}}{qd} = -k_d n_{CT} + \beta_0 n_{CS}^2, \qquad (1c)$$

where $q$ is the elementary charge and $t$ is the time. The density of separate charge carriers is represented by $n_{CS}$, where $n_{CS}^2$ corresponds to the spatial average of the product between the electron and hole densities across the active layer. Similarly, $n_S$ and $n_{CT}$ are the spatially averaged densities of excitons in the acceptor and CT states, respectively. A complete list of symbol definitions is provided in the Supporting Information.

Accordingly, excitons in the acceptor are photogenerated with a (spatially averaged) generation rate $G_S^{(A)} = G_{S,0}^{(A)} + \Delta G_{S,D}^{(A)}$, either due to direct excitations from the ground state ($G_{S,0}^{(A)}$) or via Förster resonance energy transfer (FRET) from the donor ($G_{S,D}^{(A)}$).[27] The excitons can then either undergo hole transfer to the donor and form CT states with the rate constant $k_{ht}$ or recombine with the rate constant $k_S$. The associated exciton-to-CT conversion efficiency for excitons in the acceptor is given by

$$P_S = \frac{k_{ht}}{k_S + k_{ht}}. \qquad (2)$$

Apart from being populated via excitons from the acceptor, CT states are also generated directly via dipolar transitions from the ground state, with the rate $G_{CT}$, and repopulated via bimolecular charge-carrier encounter described by the rate coefficient $\beta_0$ – noting that organic semiconductors have low permittivity which results in the formation of a CT states with near unity probability if an electron and a hole encounter. CT states are also generated from excitons in the donor with a net rate $\Delta G_{\text{eff}}^{(D)}$. In this case, we expect $\Delta G_{\text{eff}}^{(D)} = P_S^* G_S^{(D)}$, where $G_S^{(D)}$ is the generation rate of photo-induced excitons in the donor, while $P_S^* =$ $k_{et}/(k_{et} + k_S^* + k_{\mathrm{FRET}})$ is the quantum efficiency for excitons in the donor to dissociate into CT states; $k_{et}$ is the electron transfer rate to the acceptor, $k_S^*$ the recombination rate for excitons in the donor, and $k_{\mathrm{FRET}}$ the exciton FRET rate to the acceptor.

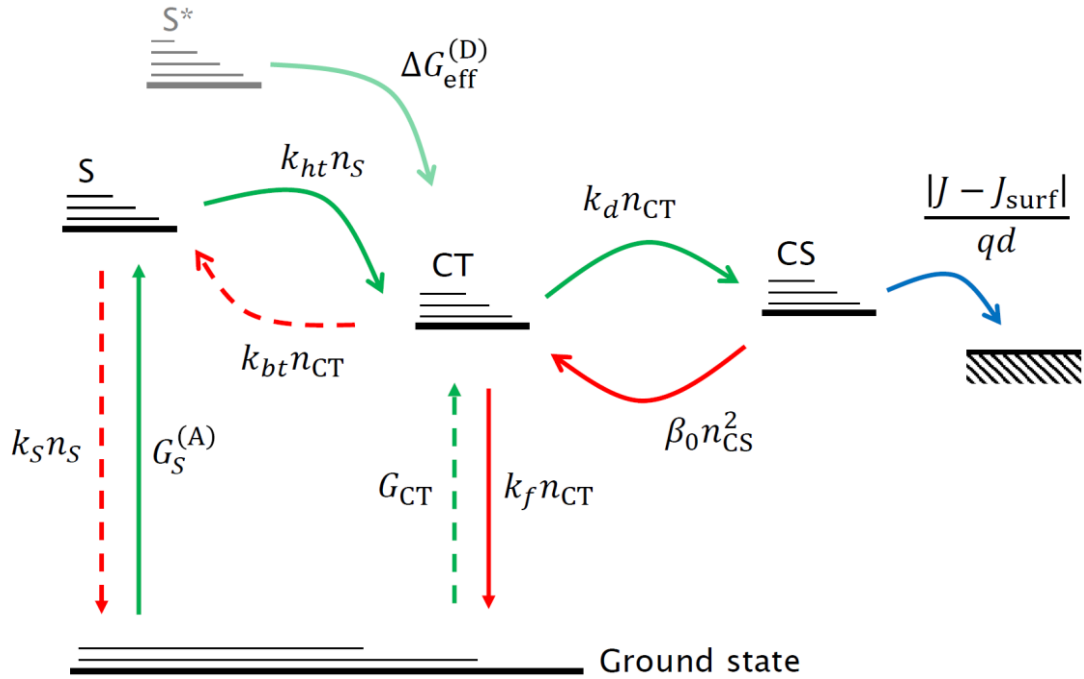

**Figure 1.** Schematic energy level diagram showing the relevant kinetics processes occurring between excitons (S) in the acceptor, charge-transfer states (CT), and free charge carriers (CS). For an active layer of thickness $d$, free charge carriers are finally collected at the electrodes with the rate $|J - J_{\mathrm{surf}}|/qd$. Additionally, CT states are also generated via photo-induced excitons (S*) in the donor.

Once generated via any of the above-mentioned routes, CT states either recombine directly to the ground state with the rate constant $k_f$, undergo a transfer back into excitons with the rate constant $k_{bt}$, or dissociate into free charge carriers with the rate constant $k_d$. The CT-state dissociation efficiency into free charge carriers, describing the ratio of CT states that do not recombine, is obtained as

$$P_{\mathrm{CT}} = \frac{k_d}{k'_{bt}+k_f+k_d}, \qquad (3)$$

where $k'_{bt} = [1 - P_S]k_{bt}$ is the ultimate back-transfer rate constant for CT states to recombine via excitons in the acceptor. Finally, as CT states dissociate, separate charge carriers, constituting free electrons and holes, are generated in the active layer. Free electrons and holes can then either encounter each other again to form CT states or be extracted at the contacts. Conversely, electrons and holes can also be injected from the contacts. The net injection-extraction rate of free charge carriers is given by $(J - J_{\text{surf}})/qd$ (see Supporting Information). Here, $J$ is the total current density, while $J_{\text{surf}}$ is the surface recombination current defined by the sum of the electron current at the anode and the hole current at the cathode.[29] Based on Equations (1)-(3), we then find

$$J = J_{\text{surf}} - qdP_{\text{CT}}\left[P_S G_S^{(A)} + P_S^* G_S^{(D)} + G_{\text{CT}}\right] + q[1 - P_{\text{CT}}]\beta_0 n_{\text{CS}}^2 d, \quad (4)$$

where the sum $\tilde{G}_{\text{CT}} = P_S G_S^{(A)} + P_S^* G_S^{(D)} + G_{\text{CT}}$ represents the effective generation rate of CT states from photons. We note that the contribution to $\tilde{G}_{\text{CT}}$ from direct CT state excitations ($G_{\text{CT}}$) is generally negligible under 1-sun illumination conditions.

## 2.1. Detailed balance

In the following we assume that excitons in the acceptor, CT states, and free electron-hole pairs are relaxed in such a way that each species are in equilibrium with themselves (i.e. can be described by a separate chemical potential). Under these conditions, the corresponding densities can be generally expressed as[30]

$$n_S = N_S \exp\left(-\frac{[E_S - \mu_S]}{K_B T}\right), \quad (5a)$$

$$n_{\text{CT}} = N_{\text{CT}} \exp\left(-\frac{[E_{\text{CT}} - \mu_{\text{CT}}]}{K_B T}\right), \quad (5b)$$

$$n_{\text{CS}}^2 = N_{\text{CS}}^2 \exp\left(-\frac{[E_{\text{CS}} - \mu_{\text{CS}}]}{K_B T}\right), \quad (5c)$$

where $K_B$ is the Boltzmann constant and $T$ is the temperature. Here $\mu_{\text{CS}}$ represents the quasi-Fermi level splitting of separate electrons and holes, $\mu_{\text{CT}}$ is the chemical potential of CT states, while $\mu_S$ is the chemical potential of the excitons in the acceptor. Furthermore $N_{\text{CS}}$, $N_{\text{CT}}$ and $N_S$ is the associated (effective) density of available states for free charge carriers, CT states, and excitons in the acceptor, respectively. Finally, $E_{\text{CS}}$ is the effective D-A energy gap for free electrons and holes.

At thermal equilibrium (i.e., in the dark at zero bias), defined by $\mu_S = \mu_{CT} = \mu_{CS} = 0$, no net currents of electrons and holes are present and $J = J_{\text{surf}} = 0$ applies. Under these conditions, detailed balance dictates all transitions to be exactly balanced by their respective inverse processes. Accordingly, we expect the CT state dissociation rate to be exactly balanced by the charge-carrier encounter rate ($k_d n_{CT,eq} = \beta_0 n_{CS,eq}^2$), while the exciton-to-CT rate is balanced by CT-to-exciton back transfer rate (i.e. $k_{ht} n_{S,eq} = k_{bt} n_{CT,eq}$ and $\Delta G_{\text{eff}}^{(D)} = 0$); hence, in conjunction with Eq. (5) we find:

$$k_d = \frac{\beta_0 N_{CS}^2}{N_{CT}} \exp\left(-\frac{\Delta E_{CT/CS}}{K_B T}\right), \tag{6}$$

$$k'_{bt} = \frac{P_S k_S N_S}{N_{CT}} \exp\left(-\frac{[E_S - E_{CT}]}{K_B T}\right), \tag{7}$$

where $\Delta E_{CT/CS} \equiv E_{CS} - E_{CT}$ and noting that $(1 - P_S) k_{ht} \equiv P_S k_S$. Here, $\Delta E_{CT/CS}$ corresponds to the (effective) CT binding energy (which generally includes a possible electric field dependence).[31-33] Based on Equations (6) and (7), Eq. (3) can be then re-expressed as

$$P_{CT} = \left[1 + \frac{k_f N_{CT}}{\beta_0 N_{CS}^2} \exp\left(\frac{\Delta E_{CS/CT}}{K_B T}\right) + \frac{P_S k_S N_S}{\beta_0 N_{CS}^2} \exp\left(\frac{\Delta E_{CT/CS} + E_{CT} - E_S}{K_B T}\right)\right]^{-1}, \tag{8}$$

being only governed by recombination rate constants and relative energetics of the different states.

A similar detailed balance is also expected for transitions between the ground state and the CT (exciton) states at thermal equilibrium: $G_{CT,eq} = k_f n_{CT,eq}$ ($G_{S,eq}^{(A)} = k_S n_{S,eq}$), where $G_{CT,eq}$ ($G_{S,eq}^{(A)}$) is the thermal generation rate of CT states (excitons). Note that, for excitons, we have assumed that the exciton energy of the donor is considerably higher than that of the acceptor such that exciton generation rate in the donor at thermal equilibrium, $G_{S,eq}^{(D)}$, is negligibly small, $G_{S,eq}^{(D)} \ll G_{S,eq}^{(A)}$. Then, after making use of Eq. (5), we find $k_f = (G_{CT,eq}/N_{CT}) \exp(E_{CT}/K_B T)$, while $k_S = (G_{S,eq}^{(A)}/N_S) \exp(E_S/K_B T)$. In accordance with the principle of reciprocity between absorption and emission,[34] we further expect $G_{CT,eq} = (\text{EQE}_{EL,CT}^{-1}/d) \int_0^\infty \eta_{\alpha,CT}(E) \phi_{BB}(E) dE$ and $G_{S,eq}^{(A)} = (\text{EQE}_{EL,S}^{-1}/d) \int_0^\infty \eta_{\alpha,S}(E) \phi_{BB}(E) dE$, allowing for the rate constants $k_f$ and $k_S$ to be determined. Here, $\eta_{\alpha,CT}$ and $\eta_{\alpha,S}$ is the quantum efficiencies for an incoming photon to excite a CT state and an exciton (either in donor or acceptor), respectively, which generally include optical interference effects as well.[35] Finally, $\text{EQE}_{\text{EL,CT}}$ ($\text{EQE}_{\text{EL,S}}$) is the electroluminescence quantum efficiency of CT states (excitons), whereas $\phi_{\text{BB}}(E) \approx (2\pi E^2/h^3c^2)\exp(-E/K_BT)$ is the black-body photon spectrum of the environment ($h$ is the Planck constant and $c$ is the speed of light).

It should be emphasized that the above detailed balance analysis generally only applies when free charge carriers, CT states, and excitons in the acceptor are relaxed [Eq. (5)]. If this condition is not met under steady-state operation (hot states), then the related rate constants may differ from the ones at thermal equilibrium.

## 2.2. Steady-state operating conditions

Under illumination, we generally have $G_{\text{CT}} \gg G_{\text{CT,eq}}$ and $G_S^{(A)} \gg G_{S,\text{eq}}^{(A)}$. Provided that excitons in the acceptor, CT states, and free charge carriers remain relaxed under steady-state conditions, simplified expressions for the current density can be derived. Based on the above detailed balance considerations, the concomitant steady-state current density of charge carriers (Eq. (4)) is obtained as

$$J = -J_{\text{gen}} + J_0 \exp\left(\frac{\mu_{\text{CS}}}{K_BT}\right) + J_{\text{surf}}, \tag{9}$$

where $J_{\text{gen}} = qdP_{\text{CT}}\tilde{G}_{\text{CT}}$ is the generation current density in the active layer. The second term on the right-hand side of Eq. (9) represents the bulk recombination current density between free charge carriers, with

$$J_0 = qdP_{\text{CT}}\left[P_Sk_SN_S \exp\left(-\frac{E_S}{K_BT}\right) + k_fN_{\text{CT}} \exp\left(-\frac{E_{\text{CT}}}{K_BT}\right)\right]. \tag{10}$$

being the associated dark saturation current density.

In the dark, the (thermal) generation current density is governed by its thermal equilibrium value $J_{\text{gen}} = J_0$, while $\mu_{\text{CS}} \approx qV$, where $V$ is the applied voltage. Assuming surface recombination to be absent, the dark current density then takes the familiar form $J_{\text{dark}} = J_0[\exp(qV/K_BT) - 1]$. Under illumination (or high injection levels), however, a deviation between the applied voltage and the quasi-Fermi level splitting of charge carriers in the bulk is generally expected for $J \neq 0$ in low-mobility (transport-limited) systems, such as organic solar cells, resulting in $\mu_{\text{CS}} \neq qV$.[36,37] The accompanying recombination rates in the bulk, taking place either via CT states at the D-A interface ($R_{CT} = k_f n_{CT}$) or via excitons in the acceptor ($R_S = k_S n_S$), are given in the Supporting Information.

## 3. Results and Discussion

To demonstrate the physical meaning and how the energetics and kinetics between excitons (in the acceptor), CT states and free charge carriers collectively determine the device performance of organic solar cells, we next conduct analytical simulations based on the developed theoretical framework. In the following, we consider the case with ideal contacts (no surface recombination), corresponding to $J_{surf} = 0$. In the simulations, we assume that the CT state generation under illumination is dominated by hole transfer from the acceptor ($\tilde{G}_{CT} = P_S G_S^{(A)}$).[27,38] However, we stress that the theoretical analysis (presented below) does not rely on this assumption and remains agnostic about the origin of $\tilde{G}_{CT}$. Furthermore, we assume $N_S = N_{CS} = 3 \times 10^{20}$ cm$^{-3}$, being close to typical number densities of acceptor molecules in the blend,[25] and $N_{CT} = 10^{18}$ cm$^{-3}$, assuming a ratio of 0.33% between interfacial states and excitonic states.[21] Finally, unless otherwise stated, we use the following default values in the simulations: $G_S^{(A)} = 1.8 \times 10^{22}$ cm$^{-3}$s$^{-1}$, $P_S = 0.99$, $E_S = 1.4$ eV, and $k_f = k_S = 10^{10}$ s$^{-1}$, while $\beta_0 = 5 \times 10^{-10}$ cm$^3$s$^{-1}$, $d = 100$ nm, and $T = 300$ K.

### 3.1. Photocurrent generation of free charge carriers

The generation current density $J_{gen}$ in Eq. (9), obtained when all photogenerated charge carriers are extracted from the active layer, can be expressed as $J_{gen} = q G_{CS} d$, where $G_{CS}$ represents the effective (average) generation rate of free charge carriers. Assuming exciton generation in the acceptor to be the dominant excitation mechanism, we expect $G_{CS} = P_{CGY} \times G_S^{(A)}$, where $P_{CGY} = P_{CT} P_S$ is the associated charge generation yield (CGY) i.e., the ratio of excitons, generated in the acceptor, that result in free charge carriers. In accordance with Eq. (8), CGY can equivalently be written as

$$P_{CGY} = \frac{P_S}{1 + \frac{k_f N_{CT}}{\beta_0 N_{CS}^2} \exp\left(\frac{\Delta E_{CS/CT}}{K_B T}\right)\left[1 + \frac{P_S k_S N_S}{k_f N_{CT}} \exp\left(\frac{E_{CT} - E_S}{K_B T}\right)\right]}. \quad (11)$$

We note that the charge generation yield is equal to the internal quantum efficiency, assuming a charge collection efficiency of 100% (at short-circuit and low light intensities).

In **Figure 2a**, the corresponding CGY is shown as a function of exciton-CT offset $E_S - E_{CT}$ at different CT-state binding energies $\Delta E_{CT/CS}$. The CGY generally follows a sigmoidal shape with the offset. We note that this is consistent with previous experimental observations.[23,27] At large offsets ($E_S \gg E_{CT}$), the CGY saturates and simplifies to $P_{CGY,sat} = P_S \times \left[1 + \frac{k_f N_{CT}}{\beta_0 N_{CS}^2} \exp\left(\frac{\Delta E_{CT/CS}}{K_B T}\right)\right]^{-1}$. Under these conditions, the CT-to-exciton back transfer is negligible ($k'_{bt} \ll k_d$) and CGY is independent of the offset. Instead, this regime is entirely governed by the competition between the CT-state recombination and dissociation rates, with CGY strongly decreasing with increasing binding energy $\Delta E_{CT/CS}$. However, this reduction can be partly compensated for by reducing the prefactor $k_f N_{CT}/\beta_0 N_{CS}^2$. This is demonstrated in **Figure 2b**, showing the effect of different $k_S$ and $k_f$. Indeed, by reducing (increasing) $k_f$, the $P_{CGY,sat}$ can be increased (reduced). A similar effect is also expected when changing the density of (available) states for charge carriers relative to CT states: reducing $N_{CT}/N_{CS}^2$ results in an increased (entropic) driving force for charge generation.

As the energetic offset between excitons and CT states is reduced, in turn, the CGY rapidly deteriorates. This is caused by an increased CT-to-exciton back transfer rate: as $k'_{bt} \gg k_d$ (low offsets), CT states are more likely to undergo back-transfer to excitons (in the acceptor) than dissociating into free charge carriers, leading to a drastic reduction in $P_{CT}$ and hence $P_{CGY}$. The critical offset, at which $k'_{bt} = k_d$, is given by $\Delta_{CGY} = \Delta E_{CT/CS} + K_B T \ln[P_S k_S N_S/\beta_0 N_{CS}^2]$. Accordingly, to avoid losses caused by back-transfer we must have $E_S - E_{CT} > \Delta_{CGY}$. Apart from increasing the offset, this can be alternatively realized by reducing the binding energy $\Delta E_{CT/CS}$ and/or decreasing $k_S$, as demonstrated in Fig. 2(a) and (b), respectively. Interestingly, an increase in $\beta_0$ generally results in both an increased $P_{CGY,sat}$ and a decreased $\Delta_{CGY}$. This can be traced back to a corresponding improvement in $k_d$, which is directly related to $\beta_0$ through detailed balance as per Eq. (6). Hence, enhancing the charge encounter rate (which depends on the charge carrier mobilities) is generally beneficial for the charge generation yield. These findings clearly demonstrates that the charge generation yield critically depends on both the relative energetics and kinetics between the different species.

## 3.2. Bimolecular recombination between free charge carriers

To avoid charge collection losses in the Fill Factor (FF) and the $J_{SC}$, a minimal charge-carrier recombination current is desired. The associated effective bimolecular recombination rate of free charge carriers in the bulk is given by $R_{CS} = \beta n_{CS}^2$, where $\beta = [1 - P_{CT}]\beta_0$ is the bimolecular charge-carrier recombination coefficient. Based on Eq. (8), we then obtain

$$\beta = \left[\frac{1}{\beta_0} + \frac{1}{\beta_{CT}+\beta_S}\right]^{-1}, \tag{12}$$

where $\beta_{CT} \equiv \beta_0 k_f/k_d = (k_f N_{CT}/N_{CS}^2) \exp\left(\frac{\Delta E_{CT/CS}}{K_B T}\right)$ may be interpreted as an effective bimolecular recombination coefficient for charge carriers to recombine via CT states, while $\beta_S \equiv \beta_0 k'_{bt}/k_d = (P_S k_S N_S/N_{CS}^2) \exp\left(-\frac{[E_S - E_{CS}]}{K_B T}\right)$ is a corresponding effective bimolecular recombination coefficient for charge carriers to ultimately recombine via excitons. **Figure 2c** and **Figure 2d** show $\beta/\beta_0$ as a function of $E_S - E_{CT}$ at different $\Delta E_{CT/CS}$ and rate constants $k_S$ and $k_f$. As expected, $\beta/\beta_0$ displays a behaviour that is nearly a mirror-image of $P_{CGY}$ with respect to the inflection point, where a low (high) CGY translates into a high (low) $\beta/\beta_0$. In general, three separate limiting regimes can be identified for the recombination/generation dynamics of organics solar cells. These regimes are defined by both energetics (the exciton-CT offset $E_S - E_{CT}$ and the CT state binding energy) and the kinetics (CT state and exciton recombination rates). Understanding these regimes is crucial for obtaining reduced recombination which is necessary in thick junction solar cells:

For $\beta/\beta_0 \ll 1$, the charge-carrier recombination rate is limited by recombination via excitons and CT states, and $\beta \approx \beta_{CT} + \beta_S$. **Regime (i)**: In this case, at large offsets ($E_S - E_{CT}$), the back-transfer rate to excitons is negligible ($k'_{bt} \ll k_f$) and the charge-carrier recombination is only limited by recombination via CT states; in this regime, $\beta$ reaches its lower limit: $\beta \to \beta_{CT}$. In this lower limit, $\beta$ is strongly dependent on the binding energy $\Delta E_{CT/CS}$ and the CT state recombination constant $k_f$, but is independent of the charge-encounter rate. CGY in this regime is near unity. **Regime (ii)**: At smaller offsets, however, $\beta$ is drastically increased (relative to $\beta_{CT}$) as $\beta \approx \beta_S$, corresponding to the ($E_S - E_{CT}$)-dependent region in Fig. 2(c) and (d). In this regime, back-transfer to excitons (in the acceptor) dominates over CT state recombination ($k'_{bt} \gg k_f$) and the charge-carrier recombination becomes instead limited by recombination via excitons. In this case, $\beta$ depends on $k_S$ and the relative energy offset between free charge carriers and excitons (in the acceptor) but is independent of the properties of the CT states and the encounter rate. *Regime (iii)*: Finally, as $\beta$ is further increased either by further decreasing the offset or increasing $\Delta E_{CT/CS}$, $k_S$, or $k_f$, $\beta$ eventually reaches its upper limit: $\beta \to \beta_0$. In this limit, the charge-carrier recombination is encounter-limited, and $\beta$ only depends on the transport properties and the morphology of D-A active layer (via $\beta_0$), with the CT states and/or excitons mostly acting as dead-ends.

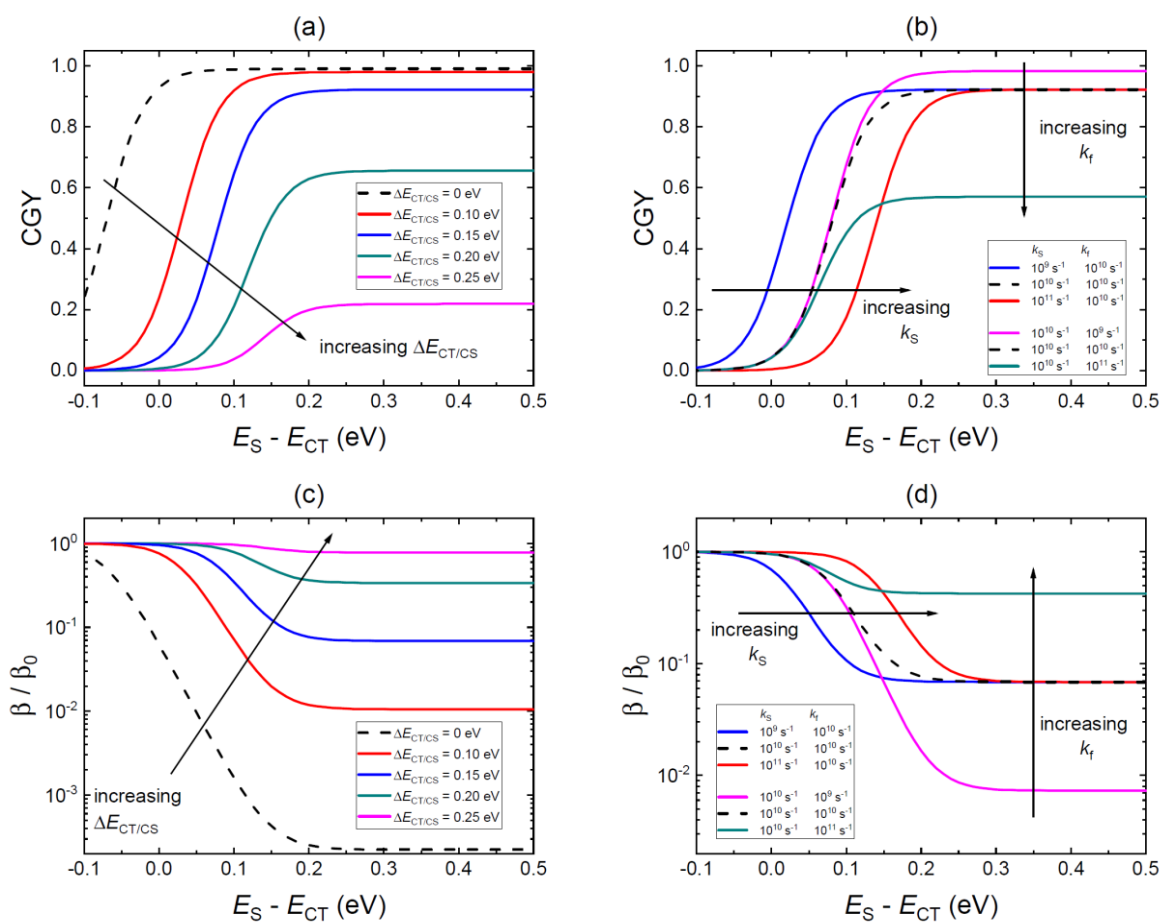

**Figure 2.** The CGY of excitons in the acceptor as function of the energetic offset, $E_S - E_{CT}$, between excitons and CT states is shown for (a) varying CT binding energy $\Delta E_{CT/CS}$, and (b) different recombination rate constants for CT states and excitons. In (c) and (d), the corresponding bimolecular charge-carrier recombination coefficient $\beta/\beta_0$, relative to the encounter rate coefficient, is shown for the cases simulated in (a) and (b), respectively.

### 3.3. The open-circuit voltage and associated photovoltage losses

While increasing the energy offset between excitons and CT states results in improved charge generation yields and lower $\beta$, this is however also expected to increase photovoltage (or $V_{OC}$) losses. Provided that surface recombination remains negligible ($J_{surf} = 0$), we expect $\mu_{CS} = qV_{OC}$ to apply at open-circuit ($J = 0$). Combining Eq. (9) and (10), we then find the following relation for the open-circuit voltage:

$$qV_{OC} = E_{CT} - K_BT \ln\left[\frac{k_f N_{CT}}{\tilde{G}_{CT}}\right] - K_BT \ln\left[1 + \frac{P_S k_S N_S}{k_f N_{CT}} \exp\left(\frac{E_{CT} - E_S}{KT}\right)\right], \qquad (13)$$

where $\tilde{G}_{CT} = P_S G_S^{(A)} + P_S^* G_S^{(D)} + G_{CT}$. The $V_{OC}$ critically depends on the interplay between excitons and CT states. Paradoxically, however, the $V_{OC}$ is ultimately independent of the energy and kinetics of free charge carriers. This is a direct consequence of the fact that the charge-carrier encounter rate is in this case exactly balanced by the dissociation rate of CT states, $k_d n_{CT} = \beta_0 n_{CS}^2$, resulting in a mutual equilibrium between free charge carriers and CT states. Indeed, further making use of Eq. (5) and (6), we find $\mu_{CT} = \mu_{CS}$, independent of $P_{CT}$. Hence, at open-circuit, free charge carriers and CT states are always in equilibrium with each other.

We note that, despite of the existing mutual equilibrium at open-circuit, the corresponding recombination rates of CT states, $R_{CT} = k_f n_{CT}$, and of free charge carriers (to the ground state), $R_{CS} = \beta n_{CS}^2$, are generally not the same. In fact, in accordance with $k_d n_{CT} = \beta_0 n_{CS}^2$, it follows that $R_{CS} = P_{CT}(k_f n_{CT} + k'_{bt} n_{CT})$. Hence, even in the limit $k'_{bt} \ll k_f$, we still have $R_{CS} = P_{CT} R_{CT}$; this is because there will always be a fraction $1 - P_{CT}$ of the generated CT states that recombine directly to the ground state without ever forming free charge carriers. Note that for $k'_{bt} \gg k_f$, in turn, most CT states formed upon charge-carrier encounter undergo back-transfer and recombine as excitons. Finally, it is important to emphasize that the equilibrium between free charge carriers and CT state is only maintained as long as $k_d n_{CT} = \beta_0 n_{CS}^2$. When $J_{surf} \neq 0$ (or $J \neq 0$), however, this condition no longer applies, and the equilibrium between CT states and free charge carriers is subsequently disrupted ($\mu_{CT} \neq \mu_{CS}$). Similarly, this condition is also violated in case of a considerable additional generation channel for charge carriers directly from hot excitons (effectively acting as a non-zero $J_{surf} < 0$ at open-circuit).

**Figure 3a** and **Figure 3b** shows the $V_{OC}$ [Eq. (13)] as a function of the energy offset between excitons and CT states at different $k_S$ and $k_f$, respectively, assuming a fixed $E_S = 1.4$ eV and $k_{ht} = 10^{12}$ s$^{-1}$. Depending on the offset, two different regimes can be identified. At large offsets ($k'_{bt} \ll k_f$), the third term on the right-hand side of Eq. (13) is negligible and the $V_{OC}$ is limited by the recombination and energy of CT states. Under these conditions, the $V_{OC}$ takes the familiar form $qV_{OC} = E_{CT} - K_B T \ln[k_f N_{CT}/\tilde{G}_{CT}]$,[3] represented by the offset dependent region in Fig. 3 (assuming a fixed $E_S$). At small offsets ($k'_{bt} \gg k_f$), on the other hand, the third term on the right-hand side of Eq. (13) eventually becomes significant, and the open-circuit voltage simplifies as

$$qV_{OC} = E_S - K_B T \ln\left[\frac{P_S k_S N_S}{\tilde{G}_{CT}}\right]. \tag{14}$$

In this regime, the recombination of free charge carriers is predominately taking place via excitons in the acceptor. Concomitantly, the $V_{OC}$ is limited by the recombination and energy of excitons (in the acceptor), becoming independent of the energetics and kinetics of CT states altogether.

At small offsets, a mutual equilibrium between excitons in the acceptor, CT states, and free charge carriers is established as $\mu_S \to \mu_{CT} = \mu_{CS}$. In general, the relationship between the chemical potentials of excitons in the acceptor and CT states at open-circuit is given by

$$\mu_S = \mu_{CT} + K_B T \ln\left(P_S + f_S\left[1 - P_S + \frac{k_f N_{CT}}{k_{ht} N_S} \exp\left(\frac{E_S - E_{CT}}{K_B T}\right)\right]\right), \tag{15}$$

where $f_S = P_S G_S^{(A)}/\tilde{G}_{CT}$. Furthermore, in accordance with the above, we have $\mu_{CT} = \mu_{CS} = qV_{OC}$ (assuming $J_{surf} = 0$). The corresponding $\mu_S$ is shown by dashed lines in Fig. 3 for the case $f_S = 1$. As expected, the difference between $\mu_S$ and $qV_{OC}$ (= $\mu_{CT}$) indeed approaches zero as the offset between excitons and CT states is reduced in this case. For a given exciton energy, this regime corresponds to the maximum chemical potential and open-circuit voltage that can be extracted from the cell. We note that for the case when CT state generation is dominated by electron transfer from the donor, $f_S \ll 1$, a deviation between $\mu_S$ and $\mu_{CT}$ is always present, independent of offset. However, provided that $P_S$ is close to unity, this deviation is generally negligible at small offsets with excitons in the acceptor and CT states effectively being in equilibrium with each other ($\mu_S \approx \mu_{CT}$).

The requirement for an (effective) equilibrium between excitons in the acceptor, CT states and free charge carriers to be established (at open-circuit) is given by $E_S - E_{CT} < \Delta_{eq}$, where $\Delta_{eq} = K_B T \ln[k_{ht} N_S / (f_S k_f N_{CT})]$ is the related critical offset (assuming $J_{surf} = 0$). As a result, at large enough offsets, when $E_S - E_{CT} > \Delta_{eq}$, the excitons and CT states are no longer in equilibrium with each other, and a considerable chemical potential difference is formed between these states. In this regime, the associated photovoltage loss is proportional to the offset and takes the form $\mu_S - \mu_{CT} \to E_S - E_{CT} - \Delta_{eq}$. Note that $\Delta_{eq}$ depends on the ratio between the CT recombination constant $k_f$ and the hole transfer rate constant $k_{ht}$, and generally increases with decreasing $k_f/k_{ht}$. Interestingly, if $k_{ht}$ is large enough, the equilibrium condition ($\mu_S \approx \mu_{CT}$) may continue to prevail beyond the exciton-dominated $V_{OC}$ regime [Eq. (14)]; see Fig. 3.

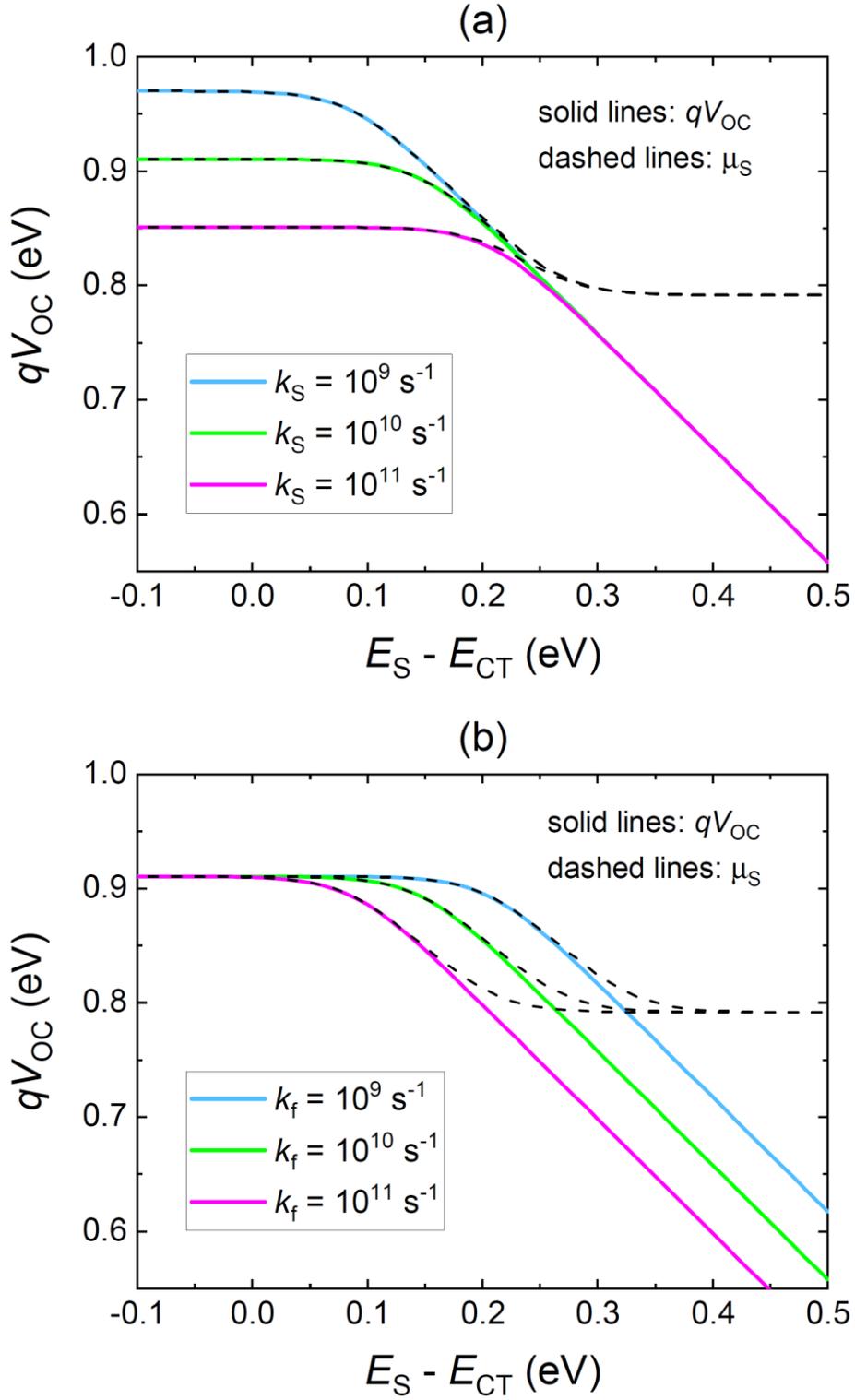

**Figure 3.** The open-circuit voltage $V_{OC}$ vs. the offset $E_S - E_{CT}$ is shown at different (a) recombination rate constants for excitons ($k_S$), and (b) CT state recombination rate constants ($k_f$). Here, $E_S = 1.4$ eV is assumed fixed, such that varying the offset is equivalent to varying $E_{CT}$. For comparison, the corresponding chemical potential for excitons $\mu_S$ is indicated by the dashed lines. Note that $qV_{OC} = \mu_{CT}$. In this case, we have assumed a fixed $k_{ht} = 10^{12}$ s$^{-1}$.

*3.4. Implications for the optimum power conversion efficiency*

Our findings are consistent with the notion that a reduction of the offset generally correlates with a concomitantly reduced photovoltage loss, resulting in an increased $V_{OC}$. This is, however, counterbalanced by a corresponding reduction of the charge generation yield and increase of the charge-carrier recombination coefficient, generally causing losses in both the $J_{SC}$ and the FF. The maximum PCE is subsequently determined by a trade-off between these two competing aspects.

To explicitly relate the generation current $J_{\text{gen}}$ ($= qG_{CS}d$) and the charge-carrier recombination coefficient $\beta$ to $J_{SC}$ and FF, a relationship between the applied voltage $V$ and the quasi-Fermi level splitting $\mu_{CS}$ is needed in Eq. (9). In principle, this requires numerical drift-diffusion simulations using $G_{CS}$, $\beta$, and the mobilities as input parameters. However, to maintain analytical tractability, yet qualitatively account for charge collection losses, we instead use an approximative analytical model developed by Neher et al.[37] Based on this so-called modified Shockey diode model, the quasi-Fermi level splitting can be simplified as

$$\mu_{CS} \approx q(V + \alpha V_{OC})/(1 + \alpha), \qquad (16)$$

where $\alpha = qd^2\sqrt{\beta G_{CS}}/(2\sqrt{u_n u_p}K_B T)$ with $u_n$ ($u_p$) being the electron (hole) mobility; here, we assume $u_n = u_p = 10^{-3}$ cm²/Vs. In accordance with Eq. (16), Eq. (9) can be then rewritten as $J = -J_{\text{gen}} + J_{\text{gen}} \exp\left(\frac{q[V-V_{OC}]}{[1+\alpha]K_B T}\right)$, allowing for $J_{SC}$ and FF to be approximated.[37]

**Figure 4** shows the corresponding PCE under 1-sun illumination as a function of the energy offset between excitons and CT states for the system under consideration. As shown, at smaller CT binding energies $\Delta E_{CT/CS}$ (Figure 4a), a lower energy offset between excitons and CT states can be afforded, simultaneously resulting in higher PCE values. We note that since the $V_{OC}$ is independent of $\Delta E_{CT/CS}$, the different curves in Figure 4a share the same $V_{OC}$ (given by green line in Figure 3). Subsequently, the PCE reduction obtained with increasing $\Delta E_{CT/CS}$ is mainly caused by an increasingly inefficient charge generation yield (reduced $k_d$), in accordance with Figure 2a. Similarly, reducing $k_S$ (and thus $k'_{bt}$) shifts the PCE maximum towards lower $E_S - E_{CT}$ allowing for higher PCE values to be reached. Conversely, by decreasing $k_f$ (Figure 4b), the PCE maximum is instead shifted towards higher energy offsets. Hence, the position of the maximum PCE is strongly sensitive to which parameter is reduced.

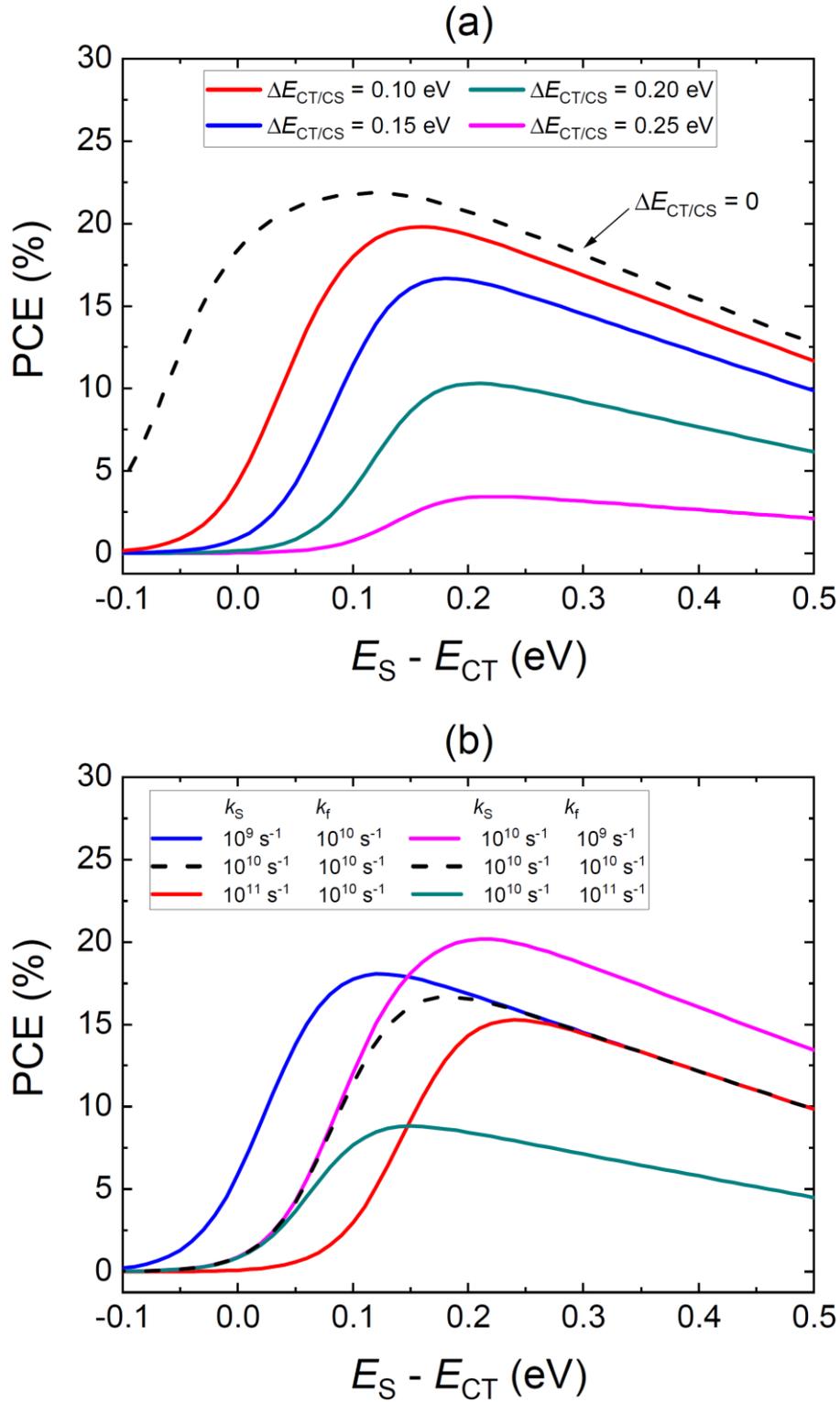

**Figure 4.** The PCE as function of the exciton-CT state offset $E_S - E_{CT}$ is shown for (a) varying CT binding energy $\Delta E_{CT/CS}$, and (b) different recombination rate constants for excitons (in the acceptor) and CT states, assuming $\Delta E_{CT/CS} = 0.15$ eV.

The exact values of the exciton and CT state rate constants are directly related to their respective radiation efficiency (or electroluminescence quantum efficiency) and the related non-radiative (NR) $V_{OC}$ losses within the active layer. For CT states, it has been suggested that these NR losses generally increases with decreasing CT state energy $E_{CT}$, in accordance with the energy-gap law.[20,21] As an increased NR loss for excitons and CT states, respectively, is expected to directly translate into an increased $k_S$ and $k_f$, this is not only expected to result in a lower open-circuit voltage, but also in a decreased charge generation yield and increased bimolecular recombination coefficient. Hence, NR losses is expected to influence the overall device performance as well as the position of the optimum energy offset.

To demonstrate this effect, we calculate the predicted PCE *versus* the optical gap, noting that the optical gap is represented by the exciton energy $E_S$ of the acceptor in this case. For the generation rate (of excitons in the acceptor), given by $G_S^{(A)} = (1/d)\int_0^\infty \eta_{\alpha,S}(E)\phi_{sun}(E)dE$, we assume step-like absorption with $\eta_{\alpha,S}(E) = 0.9$ for $E \geq E_S$, while $\eta_{\alpha,S}(E) = 0$ otherwise. Here, $\phi_{sun}$ is the AM1.5G solar spectrum, resulting in $G_S^{(A)} \approx 1.8 \times 10^{22}$ cm$^{-3}$s$^{-1}$ for $E_S = 1.4$ eV. Similarly, for CT states we assume $\eta_{\alpha,CT} = 3 \times 10^{-3}$ for $E \geq E_{CT}$, and $\eta_{\alpha,CT}(E) = 0$ otherwise. $k_S$ and $k_f$ were determined from $G_{S,eq}^{(A)}$ and $G_{CT,eq}$ via detailed balance (see Section 2.1). The radiation efficiency for CT states is given by $EQE_{EL,CT} = f_{EL}(E_{CT})$, where $f_{EL}(E_{CT})$ is a function describing the relationship between the energy and the radiation efficiency of CT states. For simplicity, we assume a similar energy dependence for the radiation efficiency of excitons (but with $E_{CT}$ replaced by $E_S$), i.e. $EQE_{EL,S} = f_{EL}(E_S)$.

The predicted PCE at different energy offsets is shown in **Figure 5a**, assuming $f_{EL}(E_{CT})$ to be given by the minimum NR loss model (for CT states) proposed by Nelson and co-workers.[21] The corresponding maximum PCE (closed symbols) is shown as a function of the exciton-CT energy offset at different binding energies $\Delta E_{CT/CS}$ in **Figure 5b**. Thus, by accounting for the energy gap dependence of $k_S$ and $k_f$, the optimum PCE is obtained at $E_S \approx 1.5$ eV (corresponding to $k_S \approx 10^7$ s$^{-1}$ and $k_f \approx 4 \times 10^7$ s$^{-1}$) for an energy offset below 0.1 eV, depending on the CT binding energy. For comparison, we also included the case with $f_{EL}(E_{CT})$ given by the lower empirical limit reported by Vandewal and co-workers[20] (as indicated by open symbols), representing the NR loss of current state-of-the-art organic solar cells. Because of the larger $k_S$ and $k_f$ in this case, a correspondingly larger PCE loss is obtained.

Nevertheless, at low binding energies, PCEs between 15-20% can be expected for offsets around 0.10-0.15 eV in this case. To enhance the PCE it is therefore crucial to reduce NR losses. An open question in this regard is whether similar NR models apply for both CT states and excitons. For example, it has been suggested that strong electronic coupling between exciton and CT states in low-offset systems cause exciton-CT-state hybridization, leading to suppressed NR recombination.[22] Answering this question is of particular importance for low-offset systems dominated by excitons in the acceptor ($k'_{bt} \gg k_f$), where minimizing the non-radiative recombination of excitons may prove to be key. Independent of the values of $k_S$ and $k_f$, however, it is clear that a small $\Delta E_{\text{CT/CS}}$ is desired to ensure maximum device performance.

## 4. Conclusions

In conclusion, we have derived an analytical framework, based on detailed balance, to clarify how the interplay between excitons, CT states, and free charge carriers influences charge generation and recombination in organic solar cells. Based on this framework, we further investigate the effect of the relative energetics and kinetics between these species on the charge generation yield, the charge-carrier recombination coefficient, and the open-circuit voltage in organic solar cell systems exhibiting small energy offsets between excitons and CT states. We find that the overall device performance is strongly dependent on the relative energetics, but critically depends on the kinetic parameters as well. Our findings highlight the importance of slow exciton recombination in low-offset systems, predicting an optimal PCE at energetic offsets of around 0.1 eV.

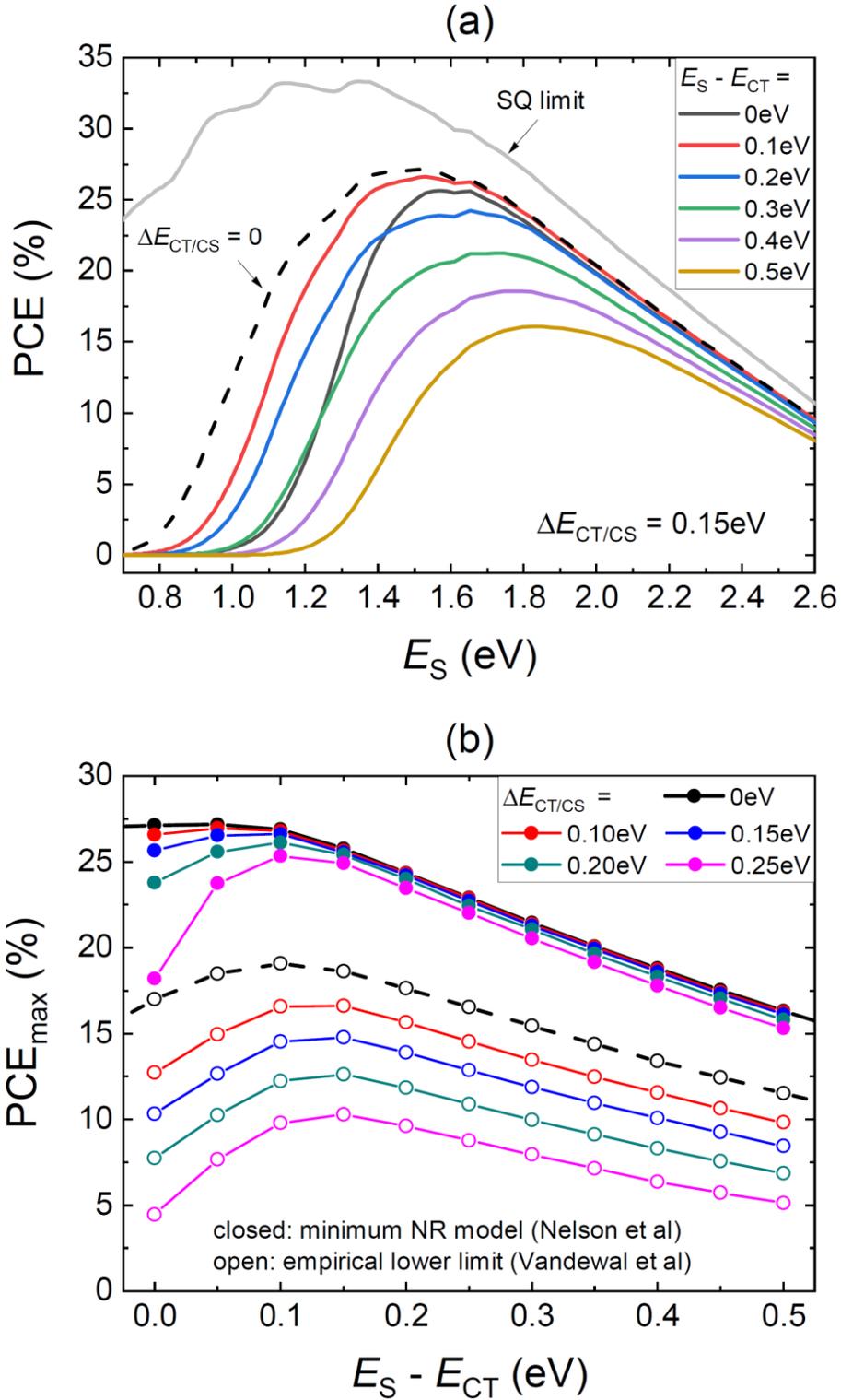

**Figure 5.** In (a), the PCE as function of the exciton energy $E_S$, corresponding to the optical gap of the active layer, is shown at different offsets $E_S - E_{CT}$, assuming a CT binding energy of $\Delta E_{CT/CS} = 0.15$eV. In (b), the corresponding maximum PCE is shown for varying $E_S - E_{CT}$ at different $\Delta E_{CT/CS}$. The ideal Shockley-Queisser (SQ) model is included for comparison in (a).

**Acknowledgements**

This work was funded by the Welsh Government's Sêr Cymru II Program through the European Regional Development Fund and Welsh European Funding Office. A.A. is a Rising Star Fellow also funded by the Welsh Government's Sêr Cymru II Program through the European Regional Development Fund, Welsh European Funding Office and Swansea University Strategic Initiative in Sustainable Advanced Materials. This work was also funded by UKRI through the EPSRC Program Grant EP/T028511/1 Application Targeted Integrated Photovoltaics. The authors thank Dieter Neher and Koen Vandewal for fruitful discussions.